\documentclass[pdflatex,sn-mathphys-num]{sn-jnl}

\usepackage{graphicx}%
\usepackage{multirow}%
\usepackage{amsmath,amssymb,amsfonts}%
\usepackage{amsthm}%
\usepackage{mathrsfs}%
\usepackage[title]{appendix}%
\usepackage{xcolor}%
\usepackage{textcomp}%
\usepackage{manyfoot}%
\usepackage{booktabs}%
\usepackage{algorithm}%
\usepackage{algorithmicx}%
\usepackage{algpseudocode}%
\usepackage{listings}%

\usepackage{siunitx}

\usepackage{svg}

\theoremstyle{thmstyleone}%
\theoremstyle{thmstyletwo}%

\theoremstyle{thmstylethree}%

\begin{document}

\title[Article Title]{On-chip, inverse-designed active wavelength division multiplexer at THz frequencies}



\author*[1]{Valerio Digiorgio}\email{vdigiorgio@phys.ethz.ch}

\author[1]{Urban Senica}
\author[1]{Paolo Micheletti}
\author[1]{Mattias Beck}
\author[1]{J\'er\^ome Faist}
\author*[1]{Giacomo Scalari}\email{gscalari@ethz.ch}

\affil[1]{Institute for Quantum Electronics, ETH Z{\"u}rich, 8093 Z{\"u}rich, Switzerland}


\abstract{The development of photonic integrated components for terahertz has become an active and growing research field. Despite its numerous applications, several challenges are still present in hardware design. We demonstrate an on-chip active wavelength division multiplexer (WDM) operating at THz frequencies. The WDM architecture is based on an inverse design topology optimization, which is applied in this case to the active quantum cascade heterostructure material embedded within a polymer in a planarized double metal cavity. Such an approach enables the fabrication of a strongly subwavelength device, with a normalized volume of only $V/\lambda^3 \simeq 0.5$.  The WDM input is integrated with a THz quantum cascade laser frequency comb, providing three broadband output ports, ranging from 2.2 THz to 3.2 THz, with $\approx$ 330 GHz bandwidth and a maximum crosstalk of -6 dB. The three ports are outcoupled via integrated broadband patch array antennas with surface emission. Such a device can be also function as a stand-alone element, unlocking complex on-chip signal processing in the THz range.}


%
%
%




\maketitle

\section{Introduction}\label{sec1}

Wavelength division multiplexing and demultiplexing are essential functions in optics and electronics, allowing parallel communication and processing of complex signals \cite{Stern_LisponOptica2015, dong2016silicon}. In the field of integrated photonics, the development of advanced optical design techniques has allowed for unprecedented integration of several components into multi-functional photonic chips for telecommunications \cite{yamada2014high, Shekhar_Bowers_NatcommRoadmap_2024} as well as for fundamental science \cite{Kim_OpticaQuantum_2020,MetcalfOnchipTeleportNatPhot_2014, Mahmudlu_QuantumsourceChip_NatPhot_2023,Griesmar_onchipSpecPhysRevRes_2021}. The THz region of the electromagnetic spectrum has recently seen a rapid evolution thanks to the advent of advanced and efficient sources \cite{lewis2014review, ding2014progress, bosco_thermoelectrically_2019,    
khalatpour2021high, khalatpour2023enhanced}, frequency combs \cite{BurghoffNatPhot2014, rosch2016chip}, and the birth of THz photonic integrated circuits \cite{SenguptaNatElec2018} that are being developed on different platforms \cite{xie2021review,Shima_Cristina_THZPIC_APLPhot_2023,Yang_Singh_TopoTHz_NatPhot2020}, including THz QCLs. The first THz wavelength division multiplexers and demultiplexers were demonstrated a few years ago but their design presents features that make on-chip integration quite challenging \cite{MittlemanWDMNatPhot2015,Ma_MIttlema_NatComms2017}. 

We recently developed an integrated THz photonic platform \cite{senica2022planarized} based on planarized waveguides, featuring an embedded quantum cascade gain medium capable of generating octave-spanning lasers \cite{Rosch_Nat.Photonics_2015_OctavespanningSemiconductorLaser}, broadband frequency combs operating both in the AM and FM regimes \cite{senica2023frequency}, and also acting as an ultrafast detector \cite{micheletti2021regenerative}. Such a platform allows ultra-broadband operation due to the double metal waveguide geometry \cite{williams2003terahertz, unterrainer2002quantum} that presents no cutoff frequency for the fundamental TM\textsubscript{00} mode. A wavelength division multiplexer (WDM) is an essential component to fully exploit the highly coherent comb sources \cite{Micheletti_solitons_SciAdv2023,senica2023frequency}, allowing signal manipulation and routing directly on-chip. 
The layout of our planarized platform comprising parallel plate metallic waveguides provides an extremely high refractive index contrast between the active region (n=3.6) and the passive material (n=1.6). Although not strictly required, this is beneficial for the exploitation of inverse design techniques \cite{molesky2018inverse} for designing compact high-performance devices, translating the pioneering work of Piggott \cite{piggott2015inverse} to 100 times longer wavelengths. A remarkable point of difference with what is already demonstrated in the near-IR is the fact that our WDM is an active device with gain, allowing us to compensate for insertion losses that can be severe at these wavelengths, as well as providing a useful knob for fine-tuning the device characteristics by introducing, for example, a microwave modulation that can be exploited in laser stabilization schemes and for spectral broadening.

\section{Results}\label{sec2}
\subsection{Inverse design and fabrication}
\begin{figure}[t!]
\centering
\includegraphics[width=0.9\textwidth]{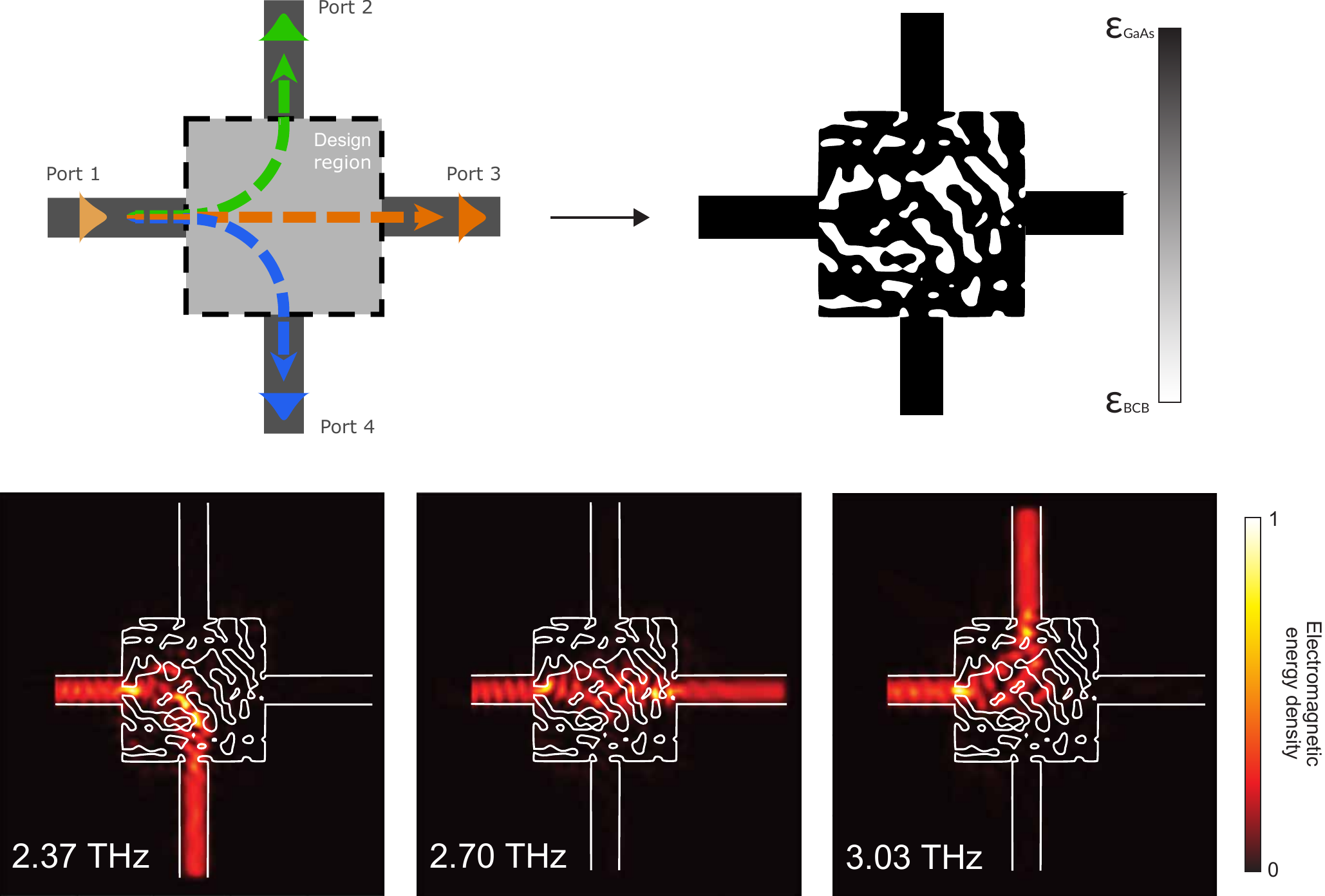}
\caption{a) Illustration of the inverse design problem from the definition of the design space to the final optimized design. b) Simulation of the power flow computed with SPINS at the central frequencies of each port.}\label{invdes_problem}
\end{figure}

The planarized platform we employ is particularly well suited for inverse design applications as the refractive index contrast between the dry-etched, high aspect-ratio active structures and the passive material sandwiched between two metals (see inset Fig.\ref{invdes_problem}) allows for a relatively simple parametrization of the devices, requiring only two discrete values for the material refractive index. The topology-optimized \cite{jensen2011topology, christiansen2021inverse} device design presented in this work has been obtained with SPINS \cite{su2020SPINSpres}, an inverse design software for running gradient-based optimization with the adjoint method \cite{lalau2013adjoint, molesky2018inverse}. The design problem requires a parametrization of the spatial material permittivity distribution $\epsilon(\text{p})$, a sequence of forward and adjoint electromagnetic simulations, and an objective function $f_{\mathrm{obj}}$, i.e., the figure of merit used to evaluate the performance of a given intermediate design. At each step of the optimization problem, the objective function and its gradient are evaluated, iteratively updating the topology until converging at the optimal design.

In particular, we use SPINS to design a THz WDM with a 200 \SI{}{\micro\metre} $\times$ 200 \SI{}{\micro\metre} compact footprint and 40 \SI{}{\micro\metre} wide input and output waveguides, routing frequencies in the 2.2 THz - 3.2 THz range to three broadband output ports. Figure \ref{invdes_problem}(a) shows the simulation region setup. The light gray square represents the design region where the permittivity is allowed to change, while the dark gray rectangles indicate the input and output waveguides.

The refractive index of GaAs (n$_{\text{GaAs}}\approx 3.6$, dispersive) and BCB (n$_{\text{BCB}}=1.57$, constant), a low-loss polymer employed for the planarization, are the upper and lower bounds set for the discrete parametrization of the permittivity distribution.
Due to the symmetric planarized waveguide structure with metallic confinement, the propagating optical waves have virtually no dependence along the vertical axis, meaning that only a 2D slice of the structure and an in-plane propagation simulation are required, resulting in a low computational load and a very efficient optimization routine. 
We defined a broadband objective function as the sum of one sub-objective per optimization frequency ($\omega_i$), maximizing the transmission to the desired port. Rejection of unwanted frequencies is not explicitly included in the figure of merit. Therefore, the objective function adopted for the presented optimization can be written as

\begin{equation}
    f_{\mathrm{obj}}(\text{p})=\sum_j\left(1-\left|\xi^{\dagger}_{i,j}\textbf{E}_j(\text{p})\right|^2\right),
\end{equation}
where $\textbf{E}_j(\text{p})$ is the electric field at frequency $\omega_j$ simulated with the parametrized permittivity distribution $\epsilon(\text{p})$ and the quantity $(\xi^{\dagger}_{i,j}\textbf{E}_j)$ represents the overlap integral used to evaluate the coupling efficiency to the desired output mode $\xi_j$ at port $i$.

The source injects the fundamental TM$_{00}$ mode at the input waveguide. The transmitted power is computed as the squared overlap integral of the simulated electric field with the TM$_{00}$ mode at the output waveguides.

Figures \ref{invdes_problem}(b,c) show the optimized design obtained after 105 iterations and the simulated power flow through the structure at the central frequency of each port.


\begin{figure}[t!]
\centering
\includegraphics[width=1\textwidth]{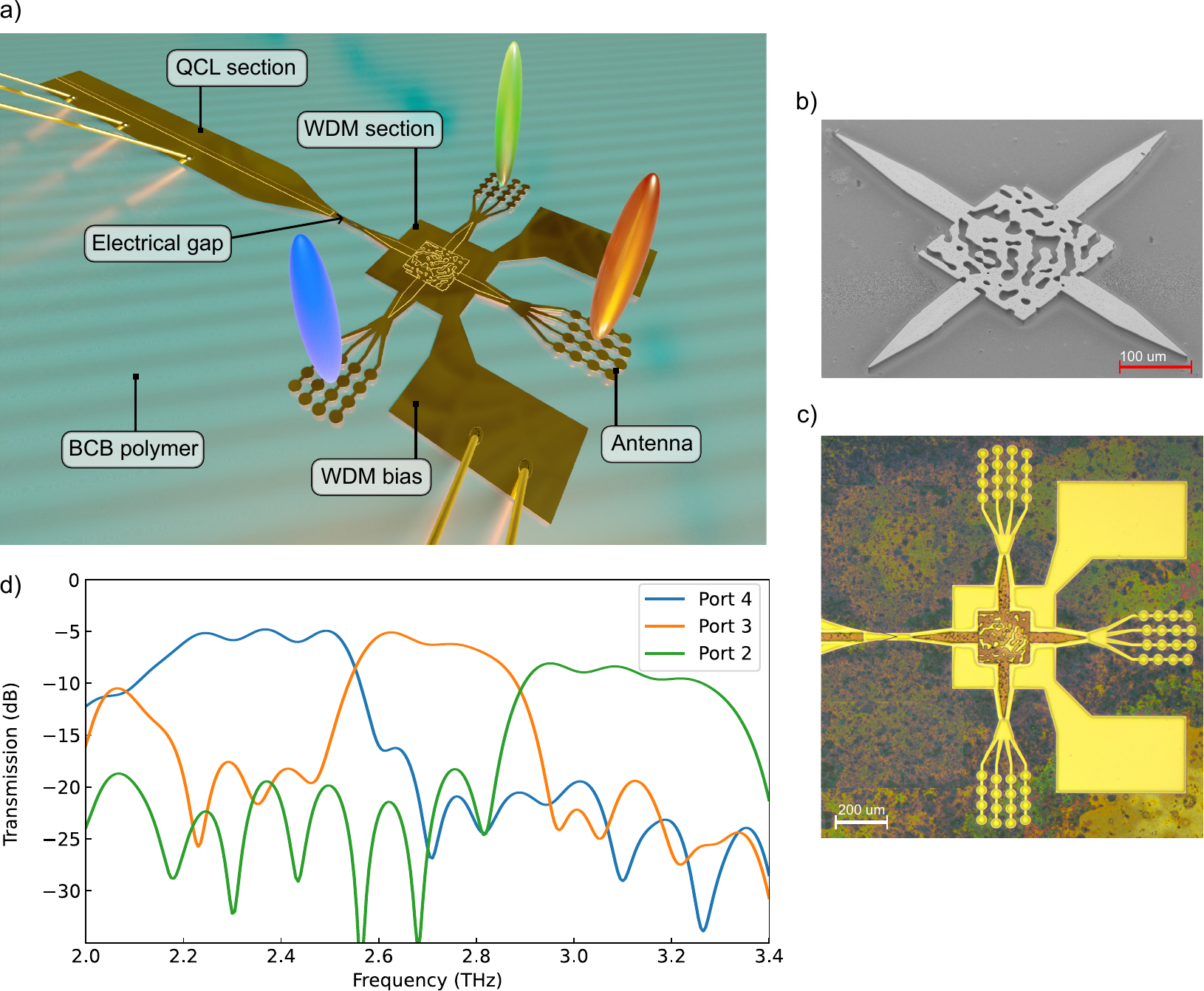}
\caption{(a) Illustration of the planarized device structure with the active material surrounded by BCB. The device combines two sections: a 2.5 mm long lasing cavity and an electrically separated WDM section with surface-emitting broadband patch array antennas. (b) SEM image of the dry-etched WDM inverse-designed region before planarization. (c) Optical microscope image of a fully fabricated device. (d) 3D CST numerical simulation results for the transmission of the QCL section output to the WDM output ports.}\label{Fabrication}
\end{figure}

The devices were fabricated following the standard procedure described in Ref. \cite{senica2022planarized}: we employed a broadband active region, similar to the one reported in Ref. \cite{forrer_photon-driven_2020}. This active region features a very broad spectral coverage (2.35-4.00 THz \cite{senica2022planarized}) as well as a low threshold current density, making it extremely suited as a source for testing the operation of our broadband, three-port WDM. 
Images of the device through different stages of fabrication are reported in Fig.\ref{Fabrication}, together with an illustration of the full structure. The lasing cavity connects to the central body of the WDM via a short passive waveguide shaped as an adiabatic tapered transition for minimizing reflections. While providing efficient optical coupling, it also allows the electrical separation of the two active sections thanks to a lithographically defined V-shaped gap in the top metallization. The three ports are outcoupled through spatially separated surface-emitting broadband patch array antennas providing excellent beam properties (both an efficient extraction and a well-defined direction) to couple the laser to a user-defined external device or detector. As presented in more detail in Refs. \cite{bosco2016, senica2023broadband}, the emission mechanism is based on an array of metallic patch elements oscillating in phase (similar to a dipole emitter array), resulting in a narrow single-lobed beam in the vertical direction. The emitted radiation exhibits a near-linear elliptical polarization along the direction of the feeding lines connecting the patches. Further optimization of the antennas would eventually enable a polarization selection of the output (together with the WDM spectral one), enable external cavity configurations \cite{hugi2010external, wysocki2005widely}, or even employ inverse design on the antenna shape for custom emission properties, such as optical vortex beams \cite{white2022inverse}. 
A broadband time-domain 3D numerical simulation of the entire device structure, comprising QCL and WDM sections (without outcoupling antennas) was performed with CST Studio Suite to predict the performances of the system (see Fig.\ref{Fabrication}(d). The results of this simulation do not include the frequency-dependent antenna response and reflectivity.

\subsection{Measurements}

\begin{figure}[b]
\centering
\includegraphics[width=0.8\textwidth]{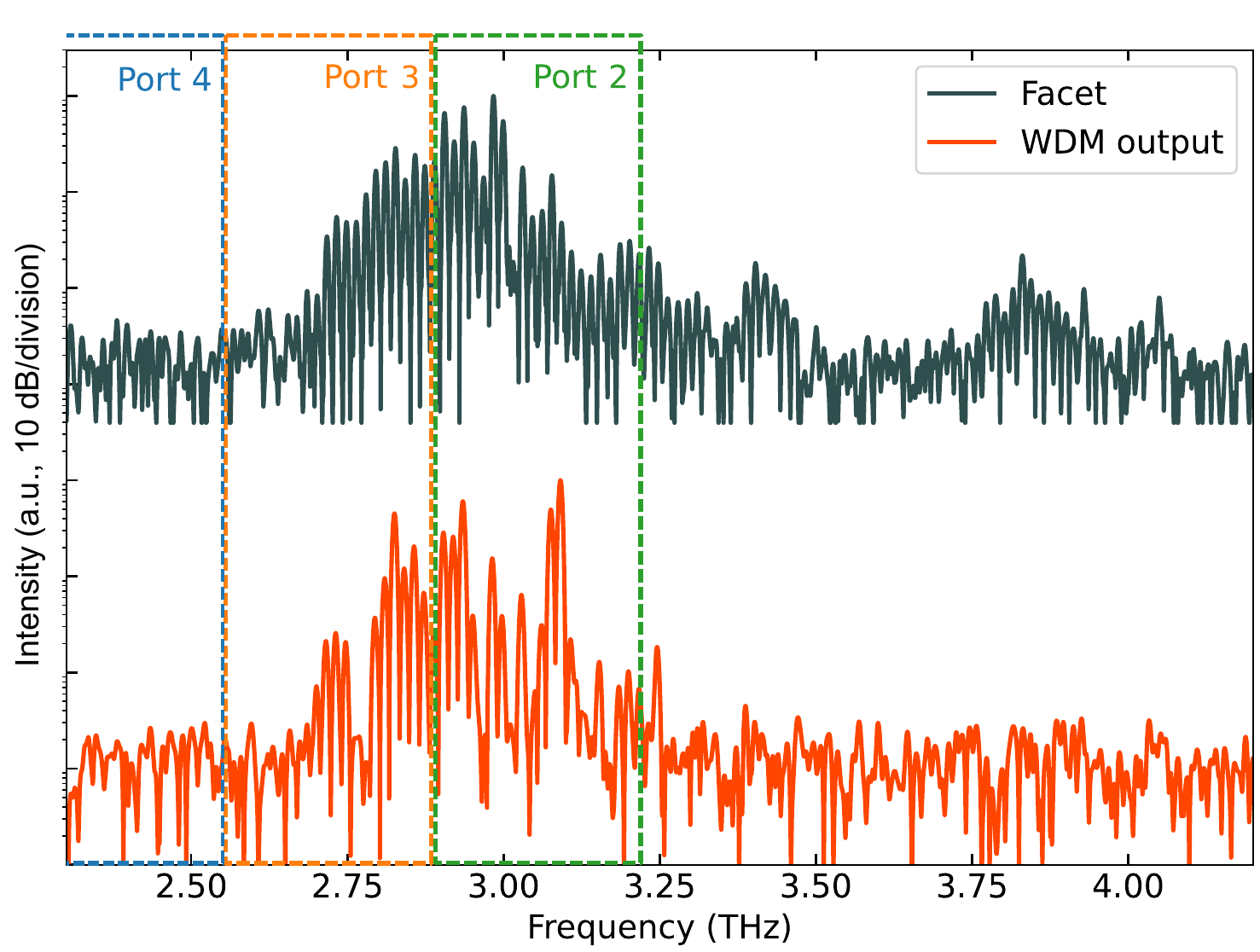}
\caption{Comparison between the spectra measured at the same operating point from the back facet of the QCL (top curve) and collectively from the three antennas of the WDM section. The curves have a vertical offset for clarity. The laser is operated at 20 K and biased at 9.4 V in continuous-wave mode.}\label{Spectrum_facet_vs_wdm}
\end{figure}

First, we test the emission of the device by collecting the signal from the ensemble of the three antennas together and comparing it with the signal emitted from the device's rear cleaved facet, where there is no spectrally selective element (see Fig.\ref{Spectrum_facet_vs_wdm}). The rear facet spectrum is much broader, covering the  2.65 - 4.00 THz bandwidth, with an FSR of 15.65 GHz matching the cavity length (2.5 mm) of the QCL section. The output of the three antennas instead covers the bandwidth  2.65 - 3.30 THz, as expected from the optimization and simulation with SPINS. Unfortunately, the measured emission spectrum is slightly blue-shifted with respect to the intentionally designed active region bandwidth, as it does not extend to lower frequencies (below 2.6 THz) that would be routed into port 4. As a consequence, we expect only some low-intensity, unfiltered broadband emission from port 4.

\begin{figure}[t]
\centering
\includegraphics[width=0.9\textwidth]{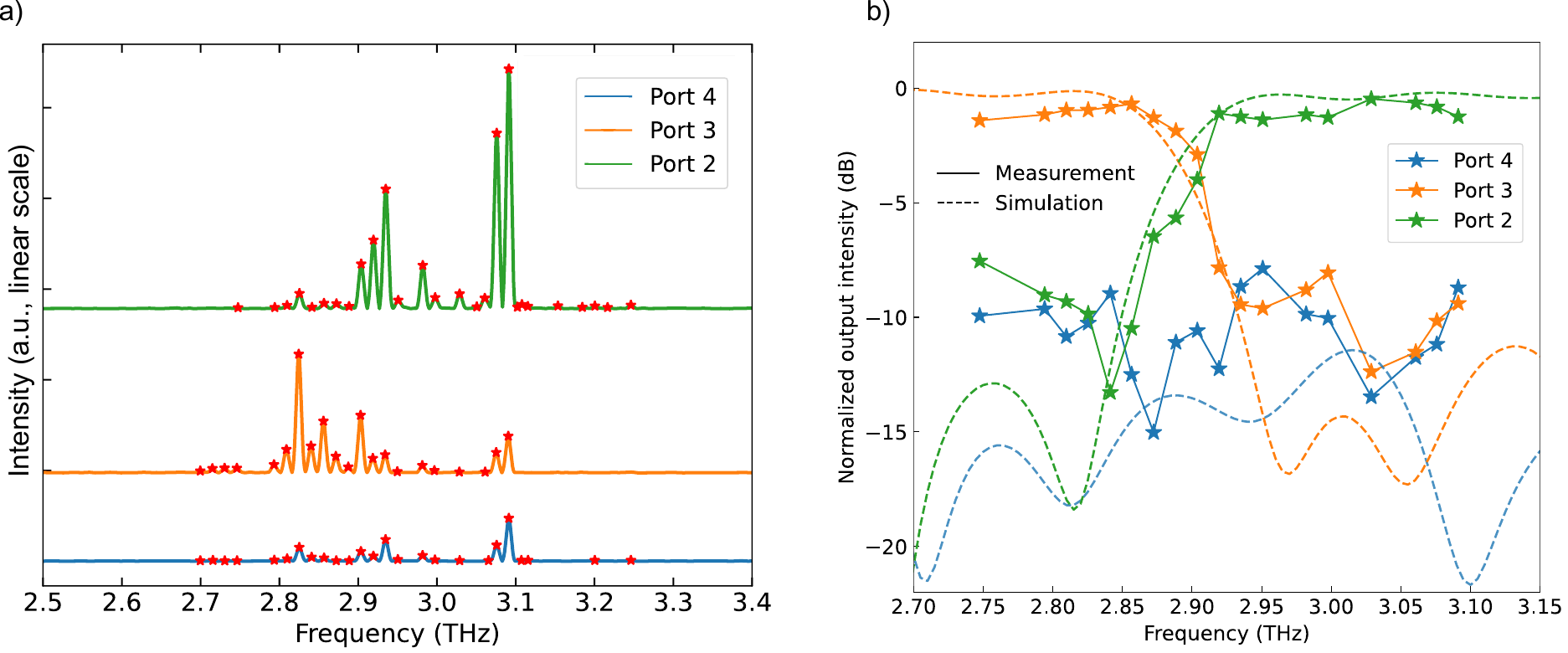}
\caption{(a) Spectral output collected individually from each port. The red stars correspond to the measured points in (b). The spectra are offset in the vertical direction for clarity without rescaling, showing the actual strength of the detected signal. (b) Normalized experimental spectral output for the three WDM ports together with the corresponding prediction from the CST 3D simulation. The measurements were done at 20 K, while the laser was operated  in continuous-wave at 9.4 V and the WDM section was biased at 8.5 V}\label{WDM_channels}
\end{figure}

\begin{figure}[b]
\centering
\includegraphics[width=0.9\textwidth]{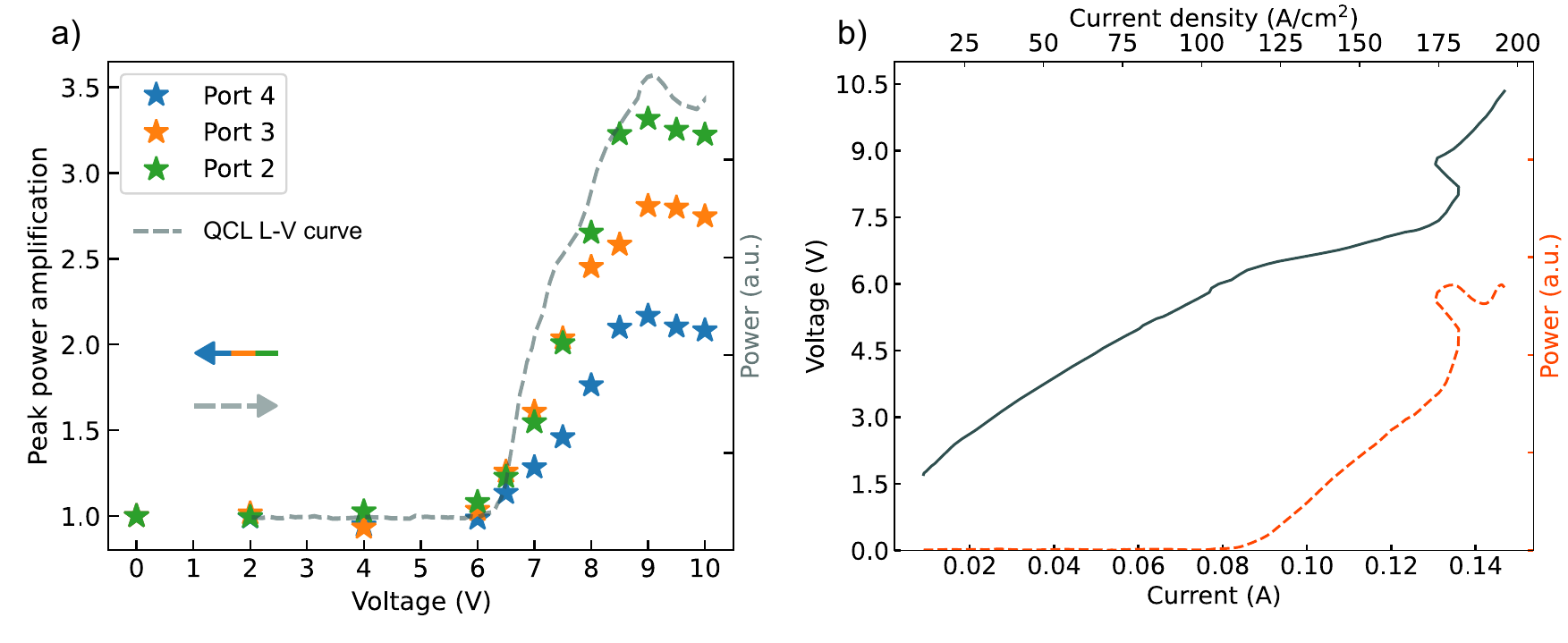}
\caption{(a) Spectrally integrated emission signal (stars) for each port as a function of the WDM applied bias. The QCL output power (dashed line) as a function of the QCL applied bias is shown for comparison of the threshold and peak output power voltages with the values for the threshold and maximum amplification of the WDM. The resemblance is clear, as the two sections are made of the same active material, featuring the same gain. (b) L-I-V curve of the laser while driven at 20 K in the micro-pulse (500 kHz, 5\% duty cycle), macro-pulse mode (30 Hz, 50\% duty cycle), with the WDM unbiased.}\label{WDM_amp}
\end{figure}

In order to evaluate the performance of the WDM experimentally, a spatial filter (a 1 mm-diameter hole within a 10x10 mm metallic plate frame) is positioned inside the cryostat in front of each antenna by means of a piezoelectric controller, allowing the collection of the signal from the individual ports.
The results of the measurements normalized to the maximum emission per port are reported in Fig.\ref{WDM_channels}. The output of ports 2 and 3 follow with good agreement the prediction, displaying a maximum crosstalk of -6 dB, in very good agreement with the calculation. Port 4, centered at 2.37 THz, shows an attenuated signal since the laser output spectral emission does not cover the targeted bandwidth, as visible from Fig.\ref{Spectrum_facet_vs_wdm}. This experimental configuration also allows us to test the amplifying characteristics of the WDM. Being electrically isolated from the laser, the WDM can be biased independently and act as an amplifier for the emitted radiation in each channel. We report in Fig.\ref{WDM_amp}(a) the integrated emitted power per port: it is clearly visible that the amplification shows a threshold at 6 V and reaches its maximum (saturates) for an applied bias of  9V, in very good agreement with what was observed from the LIV curve of the laser (Fig.\ref{WDM_amp}(b)), where the rollover point is reached at the same bias. The maximum amplification, with a factor of almost 3.5-times (5.4 dB), is achieved for port 2. This is an indication of the higher gain of the active region (biased at 9 V) at the frequencies transmitted to port 2, confirmed by the higher integrated spectral output power of this port (see Fig.\ref{WDM_channels}), followed by port 3 and finally port 4. 

\begin{figure}[htb]
\centering
\includegraphics[width=\textwidth]{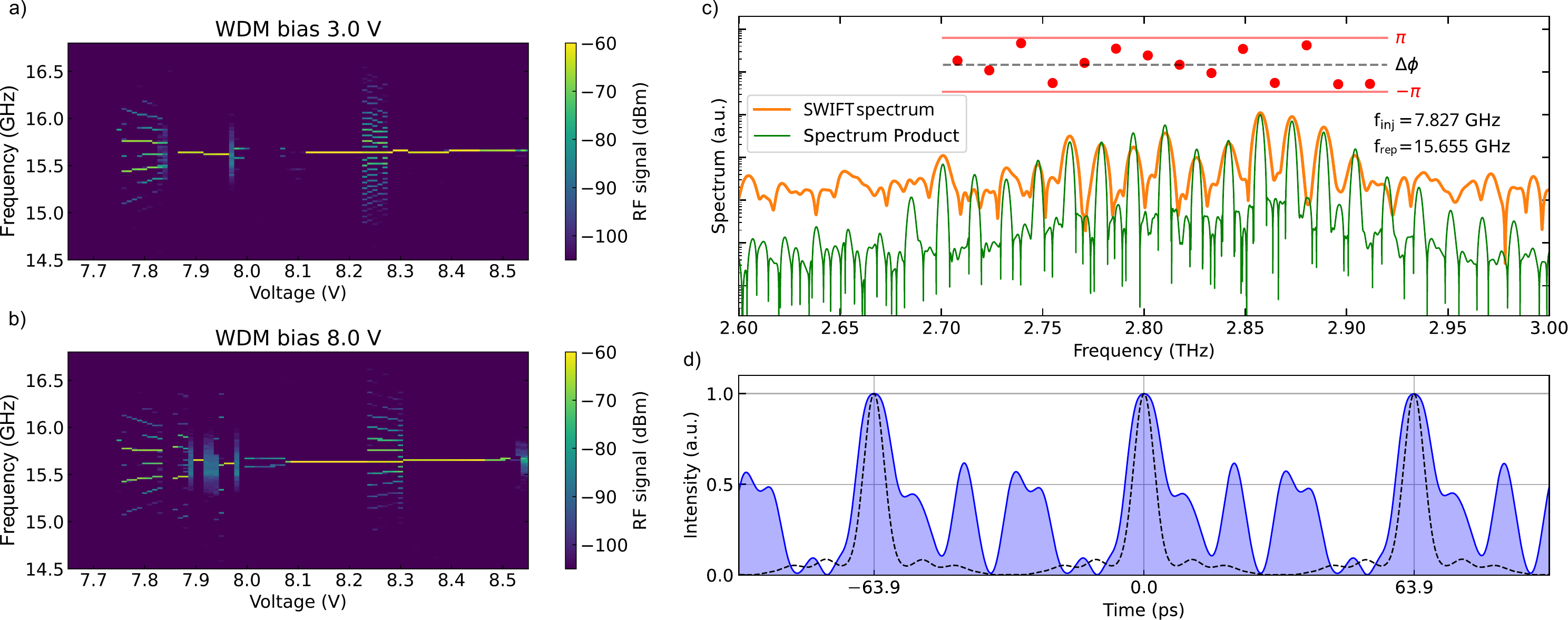}
\caption{(a,b) Electrical measurements of the beatnote of the QCL obtained at 20 K, sweeping the voltage of the laser in the two different operating regimes of the WDM: (a) without amplification, biased at 3 V; (b) with amplification, biased at 8 V.(c) Experimental SWIFT spectroscopy measurement. The laser is operated at 7.9 V (105 mA), at 20 K. The spectral product and the SWIFT spectrum overlap where the signal of the Schottky is above the noise floor. The intermodal phase differences are displayed on top for this spectral region. (d) The reconstructed temporal intensity profile (blue area) and the one corresponding to the same spectrum with Fourier-limited pulses (dashed black line) are compared.}\label{SWIFT}
\end{figure}

Additionally, we investigate the coherence of the laser emission over the frequency range where the spectral bandwidths of the laser and of the WDM overlap. While operating the QCL in free-running mode, we observe a single strong RF beatnote over a wide fraction of the laser dynamic range (see Fig.\ref{SWIFT}(a,b)). The measurements show some instability arising in a narrow voltage interval below 8 V when the WDM is operated in its amplification regime. Nevertheless, the QCL is overall unaffected by the demultiplexer section, which suggests the latter introduces weak feedback in the laser cavity. Employing a SWIFT setup described elsewhere \cite{Micheletti_solitons_SciAdv2023} and equipped with a fast Schottky detector (f$_{\text{max}}<30$ GHz),  we operate the laser under +10 dBm RF injection at a frequency of 7.87 GHz, equal to half the cavity roundtrip frequency. We observe coherent comb operation over a bandwidth of $\simeq$ 250 GHz with $f_{\text{rep}} = 2f_{\text{inj}}$= 15.65 GHz, covering channels 2 and 3. As shown in Fig.\ref{SWIFT}(d), the time reconstruction of the QCL output after the WDM splitting appears as an oscillating periodic signal. This waveform is a consequence of the mixed AM and FM nature of the frequency comb, as suggested by the reconstructed intermodal phase difference. 

We investigated as well the stability of the comb operation with respect to the WDM bias. In Fig.\ref{SWIFT}(a,b) we report the electrical beatnote signal as a function of the applied bias to the laser section for two different WDM biases $V_{WDM}=3.0$ V (\ref{SWIFT}(a)) corresponding to a lossy regime and $V_{WDM}=8.0$ V (\ref{SWIFT}(b)) , corresponding to the amplification regime. The beatnote map has some minor changes but the majority of the comb regime is preserved for any bias of the WDM section (see Supplementary Material for more measurements). This demonstrates the very good optical isolation between the two sections, allowing a refined control on the performance of the device when operating as a frequency comb. 

\section{Conclusion}

Leveraging the advanced inverse design techniques developed at telecom wavelengths and adapting them to the ultra-broadband planarized waveguide platform for THz and microwave photonics, we demonstrate an active, three-channel WDM integrated on-chip with a THz QCL comb. In this first demonstration, such a device allows the extraction of 200 GHz-wide channels with reasonably low cross-talk below -6 dB. The extracted laser modes are phase-coherent, as proved by SWIFTS measurements. This new class of devices paves the way for many applications, such as integrated THz signal processing, broadband THz spectroscopy, and coherent THz telecommunications, as recently demonstrated at telecom frequencies \cite{marin2017microresonator,shu_microcomb-driven_2022} using frequency combs. 
The key highlight of the inverse design approach is the possibility of easily optimizing for different material platforms. Additionally, the planarized double metal configuration could allow for versatile integration with various devices, by using antennas and facilitating signal transfer for next-generation integrated photonic systems.

\backmatter

\bmhead{Supplementary information}
More details on the fabrication of the device, the implementation of the inverse design approach, the electromagnetic simulations of the structure can be found in the supplementary material. Additionally, experimental methods are presented and detailed free-running measurements of the QCL comb properties are also shown and discussed.

\bmhead{Acknowledgements}


Financial support from H2020 European Research Council Consolidator Grant (724344) (CHIC) and SNF projetc 200021-212735 are gratefully acknowledged. 
\bmhead{Data availability} 
The data that support the findings of this study are available from the corresponding authors upon reasonable request.
\bmhead{Author contribution}
G.S. conceived the idea. V.D. and U.S. designed the device and performed inverse design optimization and electromagnetic simulations. V.D. and P.M. fabricated the device with support from U.S. M.B. grew the active region heterostructure. V.D. performed the measurements and data analysis. V.D. and G.S. wrote the manuscript with support from U.S. G.S. and J.F. supervised the project and acquired funding.


\end{document}